\documentstyle[aps,prb,floats,psfig]{revtex}
\begin{document}
\title{Quantum Tunneling of the Order Parameter in Superconducting Nanowires}
\author{Dmitri S. Golubev and Andrei D. Zaikin}
\address{Forschungszentrum Karlsruhe, Institut f\"ur Nanotechnologie, 76021
Karlsruhe, Germany \\
I.E.Tamm Department of Theoretical Physics,
P.N.Lebedev  Physics Institute,
Leninskii pr. 53, 117924 Moscow, Russia}

\maketitle              

\begin{abstract}
Quantum tunneling of the superconducting order parameter gives rise
to the phase slippage process which controls the resistance of 
ultra-thin superconducting wires at sufficiently low
temperatures. If the quantum phase slip rate is high, superconductivity
is completely destroyed by quantum fluctuations and the wire resistance
never decreases below its normal state value. 
We present a detailed microscopic theory of quantum phase slips in 
homogeneous superconducting nanowires. Focusing our attention on relatively
short wires we evaluate the quantum tunneling rate for phase slips, both the
quasiclassical exponent and the pre-exponential factor.
In very thin and dirty metallic wires the effect is shown to be  clearly
observable even at $T \to 0$. Our results are fully  consistent with recent
experimental findings [A. Bezryadin, C.N. Lau, and M. Tinkham, Nature {\bf
404}, 971 (2000)] which provide direct evidence for the effect of quantum
phase slips.   
\end{abstract}

\section{Introduction}

It is well established that superconducting fluctuations play a very important 
role in reduced dimension. Above the critical temperature $T_{C}$ such
fluctuations yield an enhanced conductivity~\cite{almt}. Below $T_C$
fluctuations are known to destroy the long-range order in low dimensional
superconductors \cite{HMW}.  Does the latter result mean that
the resistance of such superconductors always remains finite (or even
infinite), or can it drop to zero under certain conditions?

It was first pointed out by Little \cite{Little} that quasi-one-dimensional
wires made of a superconducting material can acquire a finite resistance below
$T_C$ of a bulk material due to the mechanism of thermally activated phase slips (TAPS).
This TAPS process corresponds to local destruction of superconductivity
by thermal fluctuations. Superconducting phase $\varphi (t)$ can flip by $2\pi$
across those points of the wire where the order parameter is (temporarily)
destroyed. According to the Josephson relation $V=\dot \varphi /2e$ 
(here and below we set $\hbar =1$) such
phase slips cause a nonzero voltage drop and, hence, dissipative currents
inside the wire. A theory of this TAPS phenomenon was developed in Refs. 
\onlinecite{lamh,2}.
This theory yields a natural result, that the TAPS probability and, hence, 
resistance of a superconducting wire $R$ below $T_C$
are determined by the activation exponent
\begin{equation}
R(T) \propto \exp (-U/T), \;\;\;\; \;\;\;U \sim 
\frac{N_0\Delta^2 (T)}{2}s\xi(T), 
\label{TAPS}
\end{equation}
where $U(T)$ is the effective potential barrier for TAPS determined simply
as the superconducting condensation energy ($N_0$ is the metallic density 
of states at the Fermi energy and $\Delta (T)$ is the BCS order parameter) for
a part of the wire of a volume $s\xi$ where superconductivity is destroyed
by thermal fluctuations ($s$ is the wire cross section and $\xi (T)$ is the 
superconducting coherence length). At temperatures very close to $T_C$ eq. 
(\ref{TAPS}) yields appreciable resistivity which was indeed detected
experimentally \cite{Exp}. Close to $T_{C}$ the experimental
results~\cite{Exp} fully confirm the activation behavior of $R(T)$
predicted in eq. (\ref{TAPS}). However, as the temperature is lowered
further below $T_C$ the number of TAPS decreases
exponentially and no measurable wire resistance is predicted by the
theory~\cite{lamh,2} except in the immediate vicinity of the critical
temperature.

Experiments \cite{Exp} were done on small diameter
whiskers and thin film samples of typical diameters $\sim 5000$ angstrom.
Recent progress in nanolithographic technique allowed to fabricate samples
with much smaller diameters down to $\sim 10$ nm. In such systems one can
consider a possibility for phase slips to be created not only due to
thermal but also due to {\it quantum} fluctuations of a superconducting
order parameter. Mooij and coworkers \cite{Mooji} discussed this
possibility and attempted to observe quantum phase slips (QPS) 
experimentally.  

Later Giordano~\cite{Gio} performed experiments which clearly demonstrated
a notable resistivity of ultra-thin superconducting wires far below $T_{C}$.
There observations could not be adequately interpreted within the TAPS theory
and were attributed to QPS. 
Other groups also reported noticeable deviations from the TAPS
prediction in thin (quasi-)1D wires ~\cite{Her,Lin}. 

First theoretical studies of the QPS effects \cite{sm,Duan,Chang} were
performed within a simple approach based on the time-dependent Ginzburg-Landau
(TDGL) equations.  Later in Refs. \onlinecite{ZGOZ,ZGOZ2} a microscopic theory 
of QPS processes was developed with the aid of the imaginary time effective
action technique \cite{ogzb} which properly accounts for
non-equilibrium, dissipative and electromagnetic effects during a QPS event.
One of the main conclusions reached in Refs. \onlinecite{ZGOZ,ZGOZ2} is that the
QPS probability is considerably larger than it was predicted previously
\cite{Duan}. For ultra-thin superconducting wires with sufficiently many
impurities and with diameters in the 10 nm range this probability can already
be large enough to yield experimentally observable phenomena. Also, further
interesting effects including quantum phase transitions caused by interactions
between quantum phase slips were discussed \cite{ZGOZ,ZGOZ2}.

In spite of all these developments an unambiguous interpretation of the
results \cite{Gio} in terms of QPS could still be questioned because of possible
granularity of the samples used in these experiments. If that was indeed the
case, QPS could easily be created inside weak links
connecting neighboring grains. Also in this case superconducting
fluctuations play a very important role \cite{BD,many,HG}, 
however -- in contrast to the QPS scenario \cite{ZGOZ,ZGOZ2} -- the
superconducting order parameter {\it needs not to be destroyed} during  the
QPS event. 

Recently, Bezryadin,
Lau and Tinkham \cite{BT} developed a new technology which allowed them
to fabricate essentially {\it uniform} superconducting wires with thicknesses
down to $3\div 5$ nm. According to our theory \cite{ZGOZ,ZGOZ2} the QPS effects
should be sufficiently large in such systems to be observed in experiments.
And indeed, the authors \cite{BT} observed that several wires 
showed no sign of superconductivity even at temperatures well
below the bulk critical temperature. Moreover, at lower temperatures their 
resistance was found to {\it increase} with decreasing temperature,
i.e. these samples could even turn insulating at $T \to 0$. The authors \cite{BT}
also argued that their experimental data can be interpreted in terms of a
quantum dissipative phase transition \cite{s,sz} which was predicted 
\cite{ZGOZ2} also for ultra-thin superconducting wires in a
certain parameter range.

The results \cite{BT} are qualitatively consistent with previous experimental
findings \cite{Gio}. Both experimental works support our general
understanding of the role of QPS processes in mesoscopic superconducting wires
and call for more detailed theoretical studies of the QPS effects. In Refs.
\onlinecite{ZGOZ,ZGOZ2} an importance of collective modes \cite{ms} and QPS
interaction effects was mainly emphasized. These are particularly important for
long wires. On the other hand, for relatively short wires interaction between
different phase slips -- at least its spatially dependent part -- should not 
play any significant role. Let us now recall that the wires studied in
the experiments \cite{BT} are not only considerably thinner but also much
{\it shorter} than those investigated by Giordano \cite{Gio}. To give some
numbers, the length of the wires \cite{Gio} was typically  $40 \div 50$ $\mu$m
whereas the wires \cite{BT} were only $0.1\div 0.2$ $\mu$m long. At the same
time, the superconducting coherence length in the experiments \cite{BT}
was even shorter, $\xi \sim 7 \div 8$ nm, i.e. such samples 
can still be considered as quasi-1D superconductors.

Motivated by the experimental findings \cite{BT}, in this paper we will
present a detailed microscopic investigation of single quantum phase slips. We
will focus our attention on an accurate  evaluation
of the QPS tunneling rate rather than on the interaction effects between
different phase slips \cite{ZGOZ,ZGOZ2}. We will go beyond
the exponential accuracy and also evaluate a
pre-exponential function in the expression for the QPS rate. We will then use
our results for a direct quantitative comparison with the experimental results
\cite{BT}.  

The structure of the paper is as follows. In Section 2 we will formulate a
simple derivation of the effective action for our problem with an emphasis put
on the Ward identities. In Section 3 we will make use of our general results
and derive the action for a special case of ultra-thin superconducting wires.
We also evaluate the QPS rate within the exponential accuracy. Section 4 is
devoted to an estimate of the pre-exponent for this rate. Comparison with
experiments and brief conclusions are presented in Section 5. Some further
technical details are diverted to Appendix.

\section{The Model and Effective Action}

The starting point for our analysis is a model Hamiltonian that includes a
short range attractive BCS and a long range repulsive Coulomb interaction.
The idea is to integrate out the electronic degrees of freedom on the level of
the partition function, so that we are left with an effective theory in terms
of collective fields \cite{sz,popov,klein}. The partition function $Z$ is
conveniently expressed as a path integral over the anti-commuting electronic
fields $\bar{\psi}$, $\psi$ and the commuting gauge fields $V$ and ${\bf A}$,
with Euclidean action
\begin{eqnarray}\nonumber
	S=\int dx{\Big (}\bar{\psi}_{\sigma}
	[\partial_{\tau}-ieV+\xi({\bf \nabla}-\frac{ie}{c}{\bf A})]
	\psi_{\sigma}-\\
	-\lambda\bar{\psi}_{\uparrow}\bar{\psi}_{\downarrow}
	\psi_{\downarrow}\psi_{\uparrow}+ienV+
	[{\bf E}^{2}+{\bf B}^{2}]/8\pi{\Big )}\;.
\label{start}
\end{eqnarray}
Here $\xi({\bf \nabla})\equiv -{\bf \nabla}^{2}/2m-\mu+U(x)$ describes a single
conduction band with quadratic dispersion and also includes an arbitrary impurity potential, $\lambda$ is the BCS coupling
constant, $\sigma=\uparrow,\downarrow$ is the spin index, and $en$ denotes the
background charge density of the ions.
In our notation $dx$ denotes $d^{3}{\bf x}d\tau$ and we use units in which
$\hbar$ and $k_{B}$ are set equal to unity. The field strengths are functions
of the gauge fields through ${\bf E}=-{\bf\nabla}V+(1/c)\partial_{\tau}
{\bf A}$ and ${\bf B}={\bf\nabla}\times{\bf A}$ in the usual way for the
imaginary time formulation.

We use a Hubbard-Stratonovich transformation to decouple the BCS interaction
term and to introduce the superconducting order parameter
field $\tilde\Delta=\Delta e^{i\varphi}$ 
\begin{eqnarray}\nonumber
	\exp\left(\lambda\int dx\bar{\psi}_{\uparrow}\bar{\psi}_{\downarrow}
		\psi_{\downarrow}\psi_{\uparrow}\right)=
	\left[\int{\cal D}^{2}\tilde\Delta e^{-\frac{1}{\lambda}\int
		dx \Delta^{2}}\right]^{-1}\\
	\times\int{\cal D}^{2}\tilde\Delta e^{-\int
		dx\left(\frac{1}{\lambda}\Delta^{2}+
		\tilde\Delta\bar{\psi}_{\uparrow}\bar{\psi}_{\downarrow}+
		\tilde\Delta^* \psi_{\downarrow}\psi_{\uparrow}
	\right)}\;,
\label{hubb-strat}
\end{eqnarray}
where the first factor is for normalization and will not be important in the
following. As a result, the partition function now reads
\begin{eqnarray}
	Z=\int{\cal D}^{2}\tilde\Delta\int{\cal D}^{3}{\bf A}\int{\cal D}V{\cal D}^{2}\Psi
	e^{\left(-S_{0}-\int dx\bar{\Psi}{\cal G}^{-1}\Psi\right)}\;,
	\label{part} \\ \nonumber
	S_{0}[V,{\bf A},\Delta]=\int dx\left(\frac{{\bf E}^{2}+{\bf B}^{2}}
	{8\pi}+ienV+\frac{\Delta^{2}}{\lambda}\right)\;,
\end{eqnarray}
where the Nambu spinor notation for the electronic fields and the matrix
Green function in Nambu space
\begin{eqnarray}\nonumber
	\Psi=\left(\begin{array}{c}\psi_{\uparrow}\\
		\bar{\psi}_{\downarrow}\end{array}\right)\;,\;\;
	\bar{\Psi}=\left(\begin{array}{cc}\bar{\psi}_{\uparrow} &
		\psi_{\downarrow}\end{array}\right)\;;\;\; \\
	\tilde{\cal G}^{-1}=\left(\begin{array}{c}
	\partial_{\tau}-ieV+\xi({\bf \nabla}-\frac{ie}{c}{\bf A})
	\;\;\;\;\;\; \tilde\Delta\\ \noalign{\vskip 2 pt}
	\tilde\Delta^* \;\;\;\;\;\;
	\partial_{\tau}+ieV-\xi({\bf \nabla}+\frac{ie}{c}{\bf A}) 
	\end{array}\right).
\label{nambu}
\end{eqnarray}
has been introduced. After the Gaussian integral over the electronic degrees
of freedom, we are left with the final effective action
\begin{equation}
	S_{\rm eff}=-{\rm Tr}\ln\tilde{\cal G}^{-1}+S_{0}[V,{\bf A},\Delta].
\label{seff1}
\end{equation}
Here the trace Tr denotes both a matrix trace in Nambu space and a trace over
internal coordinates or momenta and frequencies. In the following ``tr'' is
used to denote a trace over internal coordinates only.

The gauge invariance of the theory enables us to rewrite the action (\ref{seff1}) in a different form, which is more convenient for us,
\begin{equation}
	S_{\rm eff}=-{\rm Tr}\ln{\cal G}^{-1}+S_{0}[V,{\bf A},\Delta],
\label{seff}
\end{equation}
where
\begin{equation}
{\cal G}^{-1}=\left(
\begin{array}{c}
\partial_\tau +\xi(\nabla)-ie\Phi +\frac{m{\bf v}_s^2}{2}-\frac{i}{2}
\{\nabla,{\bf v}_s\} \hskip 20pt
\Delta \\
\Delta \hskip 20pt
\partial_\tau -\xi(\nabla)+ie\Phi -\frac{m{\bf v}_s^2}{2}-\frac{i}{2}
\{\nabla,{\bf v}_s\}
\end{array}
\right),
\label{G-1}
\end{equation}
and we have introduced the gauge invariant linear combinations of the electromagnetic potentials and the phase of the order parameter
\begin{equation}
\Phi=V-\frac{\dot\varphi}{2e}, \hskip 10pt 
{\bf v}_s=\frac{1}{2m}\left(\nabla\varphi -\frac{2e}{c}{\bf A}\right).
\label{Phi}
\end{equation}
The curly brackets $\{A,B\}$ denote an anti-commutator.

\subsection{Perturbation theory}
The action (\ref{seff}) cannot be evaluated exactly. Here we will perform a perturbative 
expansion in $\Phi$ and ${\bf v}_s$. We will keep the terms up to the second order in 
these values. This perturbation theory is sufficient for nearly all practical purposes, 
because nonlinear electromagnetic effects (described by higher order terms) are known 
to be usually very small in the systems in question. Our general derivation holds for 
an arbitrary concentration and distribution of impurities as well as for arbitrary 
fluctuations of the order parameter field in space and time.

We split the inverse Green function (\ref{G-1}) into two parts
\begin{equation}
{\cal G}_0^{-1}=\left(
\begin{array}{cc}
\partial_\tau +\xi(\nabla) &
\Delta \\
\Delta &
\partial_\tau -\xi(\nabla) 
\end{array}
\right),
\end{equation}
and
\begin{equation}
{\cal G}_1^{-1}=\left(
\begin{array}{cc}
-ie\Phi +\frac{m{\bf v}_s^2}{2}-\frac{i}{2}
\{\nabla,{\bf v}_s\} & 0 \\
0 &
ie\Phi -\frac{m{\bf v}_s^2}{2}-\frac{i}{2}
\{\nabla,{\bf v}_s\}
\end{array}
\right).
\label{G1}
\end{equation}
The logarithm in the equation (\ref{seff}) can now be expanded in powers of ${\cal G}_1^{-1}$ and we get
\begin{equation}
{\rm Tr}\ln{\cal G}^{-1}={\rm Tr}\ln{\cal G}_0^{-1}+{\rm Tr}({\cal G}_0{\cal
G}_1^{-1})-\frac{1}{2}{\rm Tr}({\cal G}_0{\cal G}_1^{-1})^2. 
\label{S1}
\end{equation}
The Green function ${\cal G}_0$ has the form
\begin{equation}
{\cal G}_0=\left(
\begin{array}{cc}
G & F \\
F & \bar{G}
\end{array}
\right).
\label{G0}
\end{equation}
In eq. (\ref{G0}) we used the fact that the non-diagonal component $\Delta$ in the matrix
${\cal G}_0^{-1}$ is real. As a result we have $\bar{F}=F$, $F(x_1,x_2)=F(x_2,x_1)$ and $\bar{G}(x_1,x_2)=-G(x_2,x_1)$. 

\subsection{Ward identities}
The Green function ${\cal G}_0$ satisfies an important identity, which is easy to check:
\begin{equation}
{\cal G}_0^{-1}\chi - \chi{\cal G}_0^{-1}=\frac{\partial\chi}{\partial\tau}-
\{\nabla,\frac{\nabla\chi}{2m}\}\sigma_3,
\label{ward0}
\end{equation}
where $\chi$ is an arbitrary function of time and space, and $\sigma_3$ is one of 
the Pauli matrices. Multiplying this matrix identity by ${\cal G}$ from the
left and from the right side and taking the diagonal components of the
resulting matrix equation we get two identities: 
\begin{eqnarray} \nonumber
\chi G-G\chi = G\left(\dot\chi-\{\nabla,\frac{\nabla\chi}{2m}\}\right)G+
F\left(\dot\chi+\{\nabla,\frac{\nabla\chi}{2m}\}\right)F ,\\
\chi\bar{G}-\bar{G}\chi =
F\left(\dot\chi-\{\nabla,\frac{\nabla\chi}{2m}\}\right)F+
\bar{G}\left(\dot\chi+\{\nabla,\frac{\nabla\chi}{2m}\}\right)\bar{G} .
\label{ward} \end{eqnarray}
Below we will use these identities in order to decouple the effective action of the 
BCS superconductor and to reduce it to a transparent and convenient form.
It is important to emphasize again that these identities are valid for any 
impurity distribution and for any time and spatial dependence of the order parameter field.
It is also worth mentioning that the Ward identity (\ref{ward0}) is {\it not} the 
result of the gauge invariance of the theory. It remains valid even for uncharged particles.

The Ward identity related to the gauge invariance of our theory has a different form:
\begin{equation}
{\cal G}_0^{-1}\sigma_3\chi -\chi\sigma_3{\cal G}_0^{-1}=
\frac{\partial\chi}{\partial\tau}\sigma_3-
\{\nabla,\frac{\nabla\chi}{2m}\}-2i\sigma_2\Delta\chi .
\label{ward2}
\end{equation}
We will use this identity to transform the first order correction to the action. It is interesting, that in the absence of superconductivity the identities (\ref{ward0}) and (\ref{ward2}) are equivalent because the inverse Green function commutes with $\sigma_3$ in this case. 
For superconductors, however, these two identities are different.

\subsection{First order}
The first order correction to the effective action is 
\begin{equation}
S_1=-{\rm Tr}({\cal G}_0{\cal G}_1^{-1})=-{\rm tr}\left[\left(\frac{m{\bf v}_s^2}{2} -ie\Phi\right)(G-\bar{G}) - \frac{i}{2}\{\nabla,{\bf v}_s\}(G+\bar{G})\right].
\label{act1}
\end{equation}
With the aid of the Ward identity (\ref{ward2}) it easy to show that the phase of the order parameter drops out from the first order terms in the electromagnetic fields. The action $S_1$ can therefore be rewritten as
\begin{equation}
S_1=-{\rm tr}(m{\bf v}_s^2G) -\int dx\left(ien_e[\Delta]V+\frac{1}{c}{\bf j}_e[\Delta]{\bf A}\right).
\label{first}
\end{equation}
We note that in general the electron density $n_e[\Delta]$ and the current density ${\bf j}_e[\Delta]$ explicitly depend on the absolute value of the order parameter.

\subsection{Second order}
It is convenient to introduce the following notations:
\begin{equation}
\dot\theta=2e\Phi, \hskip 10pt {\cal L}=\{\nabla, {\bf v}_s\}.
\label{notations}
\end{equation}
In terms of these new variables the second order correction to the action reads 
\begin{eqnarray}
\nonumber
S_2=\frac{1}{2}{\rm Tr}({\cal G}_0{\cal G}_1^{-1})^2= 
-\frac{1}{8}{\rm tr}\left[
G\dot\theta G\dot\theta + \bar{G}\dot\theta \bar{G}\dot\theta -2 F\dot\theta F\dot\theta +\right. \\
\left.
G{\cal L}G{\cal L}+\bar{G}{\cal L}\bar{G}{\cal L}+
2F{\cal L}F{\cal L} +2G\dot\theta G{\cal L} -2\bar{G}\dot\theta\bar{G}{\cal L}
\right].
\label{secorder}
\end{eqnarray}
Here we used the properties of the Green function (\ref{G0}). The form (\ref{secorder}) of the second order correction is not quite convenient, because it contains the 
$\dot\theta {\cal L}$ terms. In order to separate $\dot\theta$ and ${\cal L}$
we use the Ward identities  (\ref{ward}). We write
$$
G\dot\theta G=\theta G-G\theta + G\{\nabla,\frac{\nabla\theta}{2m}\}G-
F\left(\dot\theta+\{\nabla,\frac{\nabla\theta}{2m}\}\right)F,
$$
$$
\bar{G}\dot\theta\bar{G}=\theta\bar{G}-\bar{G}\theta + \bar{G}\{\nabla,\frac{\nabla\theta}{2m}\}\bar{G}-
F\left(\dot\theta-\{\nabla,\frac{\nabla\theta}{2m}\}\right)F.
$$
Inserting these expressions into (\ref{secorder}) after some simple transformations we 
rewrite the second order contribution as follows
$$
S_2=-{\rm tr}(G({\bf v}_s \nabla\theta))-\frac{1}{8}\left[
GKGK+\bar{G}K\bar{G}K-GMGM+\bar{G}M\bar{G}M-G\dot\theta G\dot\theta\right.
$$
\begin{equation}
\left.-\bar{G}\dot\theta\bar{G}\dot\theta -2FKFK-2F\dot\theta F\dot\theta +2FMFM+4F{\cal L}F{\cal L}\right].
\label{act2}
\end{equation}
Here we have introduced
$$
K=\{\nabla, {\bf u}\},\;\;\;\;\;M=\{\nabla,\frac{\nabla\theta}{2m}\},
$$ 
$$
{\bf u}=\frac{\nabla\theta}{2m}+{\bf v}_s=\frac{e}{m}\left(\int\limits_{-\infty}^{\tau}d\tau' (\nabla V(\tau')) -
\frac{1}{c}{\bf A}\right).
$$
The values $\dot\theta$ and ${\bf v}_s$ are now almost decoupled. The terms containing both these values were transformed into the terms containing the linear combination of these values ${\bf u}$ which does not depend on the phase of the order parameter field. The 
action (\ref{act2}) can be simplified further.
We rewrite the first term of (\ref{act2}) as follows
$$
-{\rm tr}(G({\bf v}_s \nabla\theta))={\rm tr}\left[\left(
m{\bf v}_s^2+\frac{(\nabla\theta)^2}{4m}-m{\bf u}^2\right)G\right].
$$
Again we decouple $\theta$ and ${\bf v_s}$. Making use of the identities (\ref{ward})  
yet a couple of times we arrive at the final expression for the second order
contribution  to the effective action:
$$
S_2={\rm tr}(m{\bf v}_s^2G)-{\rm tr}(m{\bf u}^2G)-\frac{1}{4}{\rm tr}
(G\{\nabla,{\bf u}\}G\{\nabla,{\bf u}\})
$$
\begin{equation}
+\frac{1}{4}{\rm tr}(F\{\nabla,{\bf u}\}F\{\nabla,{\bf u}\})+
\frac{1}{2}{\rm tr}(F\dot\theta F\dot\theta)-
\frac{1}{2}{\rm tr}(F\{\nabla,{\bf v}_s\}F\{\nabla,{\bf v}_s\}).
\label{second}
\end{equation}

\subsection{Resulting action}
Combining all contributions, we get the final result \cite{ogzb}
\begin{equation}
S=S_s[\Delta,\Phi,{\bf v}_s]+S_N[\Delta,V,{\bf A}]+S_{em}[{\bf E},{\bf B}],
\label{action}
\end{equation}
where
\begin{eqnarray}
\nonumber
S_s=\int dx\left(\frac{\Delta^2}{\lambda}\right)
-{\rm Tr}\ln{\cal G}_0^{-1}[\Delta]+{\rm Tr}\ln{\cal G}_0^{-1}[\Delta=0]+ \\
\frac{1}{2}{\rm tr}(F\dot\theta F\dot\theta)-
\frac{1}{2}{\rm tr}(F\{\nabla,{\bf v}_s\}F\{\nabla,{\bf v}_s\}),
\label{Ss}
\end{eqnarray}
\begin{eqnarray}
\nonumber
S_N=\int dx\left(-ie(n_e[\Delta]-n)V-\frac{1}{c}{\bf j}_e[\Delta]{\bf A}
+\frac{m{\bf u}^2}{2}n_e[\Delta]\right)\\
-\frac{1}{4}{\rm tr}(G\{\nabla, {\bf u}\}G\{\nabla, {\bf u}\})+
\frac{1}{4}{\rm tr}(F\{\nabla, {\bf u}\}F\{\nabla, {\bf u}\}),
\label{SN}
\end{eqnarray}
\begin{equation}
S_{em}=\int dx\frac{{\bf E}^2+{\bf B}^2}{8\pi} .
\label{Sem}
\end{equation}

\section{Effective Action for Ultra-thin Wires}

\subsection{Averaging over the electromagnetic field}

The above expressions are  complicated and in general can hardly be evaluated
in a closed form. In this section we will focus our attention
specifically on the case of quasi-one-dimensional superconducting wires
and calculate the effective action performing several approximations. We will
argue that our procedure allows to evaluate the QPS action up to a numerical
prefactor of order one.

If one assumes that deviations of the amplitude of the order parameter field
from its equilibrium value are relatively small, the above effective action
can be significantly simplified.  We expand the general effective action (\ref{action}) in
powers of  $\delta\Delta(x,\tau)=\Delta(x,\tau)-\Delta$
(here $\Delta \equiv \Delta_{\rm BCS}$) up to the second order terms. The
next step is to average over the random potential of impurities \cite{ogzb}.
After that the effective action becomes translationally invariant both in space
and in time. Performing the Fourier transformation we obtain   
\begin{eqnarray}
S&=&\frac{s}{2}\int\frac{d\omega dq}{(2\pi)^2}\left\{
\frac{|A|^2}{Ls}+\frac{C|V|^2}{s}+ \chi_E\left| qV+\frac{\omega}{c}A\right|^2
+\chi_J\left| V +\frac{i\omega}{2e}\varphi\right|^2\right. 
\nonumber\\
&&
\left.+\,\frac{\chi_L}{4m^2}\left|iq\varphi +\frac{2e}{c} A \right|^2  +
\chi_A |\delta\Delta|^2 \right\}.
\label{a105}
\end{eqnarray}
Here $L$ and $C$ are respectively the inductance times unit length and 
the capacitance per unit length of the wire. The functions $\chi_E$, $\chi_J$, $\chi_L$ and 
$\chi_A$, which depend both on the frequencies and the wave vectors, are expressed 
in terms of the averaged
products of the Green functions appearing in the eqs. (\ref{Ss}), (\ref{SN})
(see Ref. \onlinecite{ogzb} for more details).  These functions 
can be evaluated analytically for most limiting cases.
For the sake of completeness some explicit expressions are presented in
Appendix.

The voltage $V$ and the vector potential $A$ enter the action in a quadratic form
and, hence, can be integrated out exactly. After that
the effective action will only depend on $\varphi$ and $\delta\Delta$. We find 
\begin{equation}
S=\frac{1}{2}\int\frac{d\omega dq}{(2\pi)^2}\left\{{\cal F}(\omega ,q)|\varphi|^2+ \chi_A |\delta\Delta|^2\right\},
\end{equation}
where
\begin{equation}
{\cal F}(\omega ,q)=
\frac{\left(\frac{\chi_J}{4e^2}\omega^2+\frac{\chi_L}{4m^2}q^2 \right)
\left(\frac{C}{sL}+\chi_E\left[ C\omega^2+\frac{q^2}{L}\right] \right)+
\frac{\chi_J\chi_L}{4m^2}\left[C\omega^2+\frac{q^2}{L}\right]}
{\left(\frac{C}{s}+\chi_J+\chi_E q^2 \right)\left(\frac{1}{sL}+\chi_E\omega^2+\frac{e^2}{m^2}\chi_L \right)
-\chi_E^2\omega^2 q^2} .
\end{equation}
The electromagnetic potentials are expressed as follows:
\begin{eqnarray}
V&=&\frac{\chi_J\left(\frac{1}{sL}+\chi_E\omega^2+\frac{e^2}{m^2}\chi_L \right)+
\frac{e^2}{m^2}\chi_E\chi_L q^2}
{\left(\frac{C}{s}+\chi_J+\chi_E q^2 \right)\left(\frac{1}{sL}+\chi_E\omega^2 +\frac{e^2}{m^2}\chi_L \right)
-\chi_E^2\omega^2 q^2} \left(\frac{-i\omega}{2e}\varphi \right),
\\
A&=&\frac{\frac{e^2}{m^2}\chi_L\left(\frac{C}{s}+\chi_J+\chi_Eq^2 \right)+\chi_E\chi_J \omega^2}
{\left(\frac{C}{s}+\chi_J+\chi_E q^2 \right)\left(\frac{1}{sL}+\chi_E\omega^2 +\frac{e^2}{m^2}\chi_L \right)
-\chi_E^2\omega^2 q^2} \left(\frac{icq}{2e}\varphi \right).
\end{eqnarray}

In most of the situations the wire inductance is not important and can be neglected.
Therefore here and below we put $L=0$. Then we get
\begin{equation}
S=\frac{s}{2}\int\frac{d\omega dq}{(2\pi)^2}\left\{
\frac{\left(\frac{\chi_J}{4e^2}\omega^2+\frac{\chi_L}{4m^2}q^2 \right)
\left(\frac{C}{s}+\chi_E q^2 \right)+
\frac{\chi_J\chi_L}{4m^2}q^2}
{\frac{C}{s}+\chi_J+\chi_E q^2} |\varphi|^2 + \chi_A |\delta\Delta|^2\right\},
\label{a106}
\end{equation}
and
\begin{eqnarray}
V&=&\frac{\chi_J}
{\frac{C}{s}+\chi_J+\chi_E q^2 } \left(\frac{-i\omega}{2e}\varphi \right),
\label{noJ}\\
A&=&0.
\end{eqnarray}
Let us note that the Josephson relation $V=\dot\varphi/2e$ is in general
not satisfied. According to eq. (\ref{noJ}) this relation may approximately
hold only in the limit $\chi_J\gg C/s+\chi_Eq^2$. Making use of the results
presented in Appendix one easily observes that in a practically important limit
of small elastic mean free paths $l$ the latter condition is obeyed only at low
frequencies and wave vectors $\omega/\Delta\ll 1$ and $Dq^2/\Delta\ll 1$, where
$D=v_Fl/3$ is the diffusion constant.

Let us now perform yet one more approximation and expand the action in powers
of $\omega$ and $q^2$. Keeping the terms of the order $q^4$ and $\omega^2q^2$
 we find $$
S=\frac{s}{2}\int\frac{d\omega dq}{(2\pi)^2}\left\{
\left(\frac{C}{s}\omega^2 +\pi\sigma\Delta q^2+ \frac{\pi^2}{8}\sigma Dq^4
+\frac{\pi\sigma}{8\Delta}\omega^2 q^2 \right)
\left|\frac{\varphi}{2e}\right|^2\right.
$$
\begin{equation}
\left.+2N_0\left(1+\frac{\omega^2}{12\Delta^2}+\frac{\pi Dq^2}{8\Delta} \right)|\delta\Delta|^2
\right\}.
\label{action11}
\end{equation}
The term $\propto\omega^4$ turns out to be equal to zero.
In (\ref{action11}) we introduced the normal state conductance of the
wire $\sigma =2e^2N_0D$.
At even smaller wave vectors, $Dq^2/2\Delta\ll 2C/\pi e^2N_0s \ll 1,$ we get
\begin{eqnarray}
S&=&\frac{1}{2}\int\frac{d\omega dq}{(2\pi)^2}
\left\{
\left( {C}\omega^2+\pi\sigma\Delta s q^2\right) 
\left|\frac{\varphi}{2e}\right|^2 
+ s\chi_A |\delta\Delta|^2\right\}.
\label{longwave}
\end{eqnarray}
Here we have assumed $C/2e^2N_0s\ll 1$. This inequality 
is usually well satisfied for sufficiently good metals, perhaps except
for the case of some specially chosen substrates. 
The form of the action suggests the existence of the plasma modes which 
can propagate along the wire. These are the so-called Mooij-Sch\"on modes \cite{ms},
the velocity of which is given by the equation:
\begin{equation}
c_0\simeq \sqrt{\frac{\pi\sigma\Delta s}{C}}.
\end{equation}

\subsection{QPS action}

One can show \cite{ZGOZ} that for very long wires 
the action (\ref{longwave}) yields a QPS solution described by a simple 
formula $\varphi(x,\tau)=-\arctan(x/c_0\tau)$. 
The long time behavior of this solution results in the logarithmic interaction
between two phase slips $(x_1,\tau_1)$ and $(x_2,\tau_2)$:
\begin{equation}
S_{\rm int}=\frac{\mu}{2}\ln\left[\frac{(x_1-x_2)^2+c_0^2(\tau_1-\tau_2)^2}{\xi^2}\right],
\label{Sint}
\end{equation}
where 
\begin{equation}
\mu=\frac{\pi}{4\alpha}\sqrt{\frac{sC}{4\pi\lambda_L^2}}.
\end{equation}
Here $\alpha\simeq 1/137$ is the fine structure constant. In short wires, 
however, the above logarithmic interaction (\ref{Sint}) does not
play an important role and can be essentially neglected. 

Let us estimate the contribution of a single phase slip to the effective 
action. First we rewrite the action (\ref{action11}) in the space-time 
domain dropping the unimportant term $\propto (Dq^2)^2$:
\begin{eqnarray}
S&=&\frac{s}{2}\int dx\, d\tau\left\{ \frac{C}{4e^2s} \left(\frac{\partial \varphi}{\partial\tau} \right)^2
+\frac{\pi\sigma\Delta}{4e^2} \left(\frac{\partial \varphi}{\partial x}\right)^2+
\frac{\pi\sigma}{32e^2\Delta}\left(\frac{\partial^2 \varphi}{\partial x\partial \tau}  \right)^2
\right\}
\nonumber\\
&&+\;
sN_0\int dx\, d\tau \left\{ \delta\Delta^2+\frac{1}{12\Delta^2}\left(\frac{\partial \delta\Delta}{\partial\tau} \right)^2
+ \frac{\pi D}{8\Delta} \left(\frac{\partial \delta\Delta}{\partial x} \right)^2 \right\}.
\label{action21}
\end{eqnarray}

Then we assume that the absolute value of the order parameter 
is equal to zero at a time $\tau=0$ and at a point $x=0$. The 
size of the QPS core is denoted as $x_0,$ and its time duration is $\tau_0$.
The amplitude of the fluctuating part of the order parameter field  
$|\delta\Delta(x,\tau)|$ can be approximately expressed as follows:
\begin{equation}
|\delta\Delta(x,\tau)|=\Delta \exp(-x^2/2x_0^2-\tau^2/2\tau_0^2) .
\label{Deltaqps}
\end{equation}

The QPS phase dependence on $x$ and $\tau$ 
should satisfy several requirements. In a short wire and outside the 
QPS core the phase $\varphi$ should not depend on the spatial 
coordinate in the zero current bias limit. On top of that, at $x=0$ 
and $\tau=0$ the phase should flip in a way to provide the change of the net 
phase difference across the wire by $2\pi$. For concreteness, let us present 
two different trial functions which obey the above requirements. For instance,
one may choose  
\begin{equation}
\varphi(x,\tau)=-\frac{\pi}{2\cosh(\tau/\tau_0)}\tanh\left(\frac{x}{x_0\tanh(\tau/\tau_0)}\right),
\label{phiqps}
\end{equation} 
or
\begin{equation}
\varphi(x,\tau)=-\frac{\pi}{2}\tanh\left(\frac{x\tau_0}{x_0\tau}\right).
\label{phiqps2}
\end{equation} 
Similar other trial functions can also be considered. 

Substituting the trial functions (\ref{Deltaqps}), (\ref{phiqps})
(or (\ref{Deltaqps}), (\ref{phiqps2})) into the action (\ref{action21}) 
one arrives at the expression
\begin{eqnarray} S(x_0,\tau_0)&=&\left[
a_1\frac{C}{e^2} +  a_2 sN_0\right]\frac{x_0}{\tau_0}+ a_3
sN_0D\Delta\frac{\tau_0}{x_0} \nonumber\\
&&+\,
a_4\frac{sN_0D}{\Delta}\frac{1}{x_0\tau_0} + a_5 sN_0\Delta^2 x_0\tau_0+
a_6 \frac{\tilde C}{e^2\tau_0},
\label{action1}
\end{eqnarray}
where $a_j$ are  numerical factors of order one which depend on the 
precise form of the trial functions,
$\tilde C=CX$ is the total capacitance of the wire and $X$ is the wire length.
Note that fictitious divergences emerging from a singular
behavior of the functions (\ref{phiqps}), (\ref{phiqps2}) at $x=x_0$ and
$\tau=\tau_0$ are eliminated since the order parameter vanishes inside the
QPS core. 

Let us first disregard capacitive effects neglecting the last term 
in eq. (\ref{action1}).
Minimizing the remaining action with respect to the core parameters $x_0$ and 
$\tau_0$ and making use of the inequality $C/e^2N_0s\ll 1$, we
obtain \begin{equation}
x_0=\left(\frac{a_3a_4}{a_2a_5}\right)^{1/4}\sqrt{\frac{D}{\Delta}},
\;\;\;\;\;\; \tau_0=\left(\frac{a_2a_4}{a_3a_5}\right)^{1/4}\frac{1}{\Delta}.
\label{xtau}
\end{equation}
These values provide the minimum for the QPS action, and we find
\begin{equation}
S_{QPS} = 2 (\sqrt{a_2a_3}+\sqrt{a_4a_5}) N_0s\sqrt{D\Delta} .
\label{S321}
\end{equation}
One can also express eq. (\ref{S321}) in the form convenient
for further comparison with experiments:
\begin{equation}
S_{QPS}=A\frac{R_q}{R}\frac{X}{\xi }.
\label{otvet}
\end{equation}
Here $A=2 (\sqrt{a_2a_3}+\sqrt{a_4a_5})/\pi,$ $R$ is the total wire resistance,
$R_q=\pi\hbar/2e^2=6.453$ k$\Omega$ is the resistance quantum and 
$\xi =\sqrt{D/\Delta}$ is the superconducting coherence length.

As it was already pointed out, the results (\ref{xtau}) and (\ref{S321}) 
hold provided the capacitive effects are small. This is the case for relatively 
short wires
\begin{equation}
X \ll \xi\frac{e^2N_0s}{C}.
\label{short}
\end{equation}
In the opposite limit the same minimization procedure of the action 
(\ref{action1}) yields
\begin{equation}
x_0 \sim \xi,\;\;\;\;\; \Delta \tau_0 \sim \sqrt{XC/\xi e^2N_0s} \gg 1.
\label{cap}
\end{equation}
The QPS action again takes the form (\ref{otvet}) with 
$A \sim \sqrt{XC/\xi e^2N_0s}$. 

For the sake of clarity, let us summarize the approximations performed in this
section. As a first step, we expanded the action derived in Section 2 up to
the second order in  $\delta\Delta(x,\tau)=\Delta(x,\tau)-\Delta$.
Obviously this approximation is sufficient everywhere except
inside the QPS core where $\Delta(x,\tau)$ is small. In these space-
and time-restricted regions one can expand already in $\Delta(x,\tau)$
again arriving at eq. (\ref{a105}) with $\delta\Delta(x,\tau)\to\Delta(x,\tau)$
and with all the $\chi$-functions defined in Appendix with $\Delta \equiv 0$.
Both expansions match smoothly at the scale of the core size
$x_0 \sim \xi$, $\tau_0 \sim 1/\Delta$. Hence, the approximation (\ref{a105})
is sufficient to obtain the correct QPS action, perhaps up to a numerical
prefactor of order one.

In order to simplify our analysis further, in (\ref{action11}) we expanded
(\ref{a105}) in powers of $\omega /\Delta$ and $Dq^2/\Delta$. Again this 
approximation is sufficient within the same accuracy. Indeed, one can
-- even without performing this expansion -- substitute the trial functions
(\ref{Deltaqps}), (\ref{phiqps}) (or (\ref{Deltaqps}), (\ref{phiqps2}))
directly into the action (\ref{a106}). If the capacitive effects
are neglected (\ref{short}), the resulting
QPS action can be represented as a function of the dimensionless parameters
$x_0/\xi$ and $\Delta \tau_0$ only. Making use of the general expressions
for the $\chi$-functions collected in Appendix and minimizing the QPS action
with respect to $x_0$ and $\tau_0$ one again arrives at the result  
(\ref{otvet}) with $A \sim 1$. If the inequality (\ref{short}) is violated,
the accuracy of our expansion in powers of $\omega /\Delta$ may only
become better (cf. eq. (\ref{cap})).
  
Finally, the particular choice of the trial functions 
(e.g. (\ref{Deltaqps}) and (\ref{phiqps})) describing the QPS event also
appears not to play any significant role as long as these trial functions
obey the general requirements formulated above. In addition to 
(\ref{Deltaqps}), (\ref{phiqps}) and (\ref{phiqps2})
we have used several other trial QPS functions. In all cases we have 
obtained $A$ within the interval $A \approx 0.8\div 2.5$. In a way our method
can be regarded as a variational procedure. Therefore, even though the exact value
of a numerical prefactor $A$ in eq. (\ref{otvet}) cannot be established within
our approach, we do not expect $A$ to deviate substantially from the above
values. 

Our last remark in this section concerns the role of dissipation. From the
form of the result (\ref{S321}) one could naively assume that
the correct QPS action could be guessed, e.g., from a simple TDGL-based
approach (or, alternatively, only from the ``condensation energy'' term
proportional to $\chi_A$) without taking into account
dissipative and electromagnetic effects. Indeed, minimization of the 
contribution $\sim |\delta \Delta |^2$ (the last three terms in eq.
(\ref{action21})) is formally sufficient to arrive at the correct estimate
$S_{QPS} \sim N_0s\sqrt{D\Delta}$. It is obvious, on the other hand, that not
only the amplitude but also the phase fluctuations of the order parameter are
important during the QPS event. If the latter fluctuations are taken into
account {\it without} including dissipative effects (this would correspond
to formally setting $\sigma \to 0$ in eq. (\ref{action21})) the estimate for
the QPS action $S_{QPS} \sim \mu$ would follow immediately. This result would be 
parametrically different from eq. (\ref{S321}). Note, however, that
within our model the dissipative effects can be ignored only for
$C/e^2N_0s \gtrsim 1$. Usually the latter condition cannot be satisfied
for metallic systems, perhaps except for some specially chosen substrates.
In the opposite -- more realistic -- limit $C/e^2N_0s \ll 1$ dissipation
plays a dominant role during the phase slip event, and the correct QPS action
{\it cannot} be obtained without  an adequate microscopic description of 
dissipative currents flowing inside the wire.

\section{Pre-exponent}

The above results allow to estimate the exponential suppression of QPS in
ultra-thin superconducting wires depending on thickness, impurity
concentration and other parameters. These results, however, are not yet
sufficient to evaluate the whole QPS rate which has the form 
\begin{equation}
\gamma_{QPS}=B\exp (-S_{QPS}).
\label{gamma}
\end{equation}
The task at hand is to provide a reliable estimate for the pre-exponential
factor $B$ in eq. (\ref{gamma}). A general strategy to be used for this
purpose is well known \cite{ABC}. One can start, e.g., from the expression for
the grand partition function of the wire   
\begin{equation}
Z=\int {\cal D} \Delta  {\cal D}\varphi \exp ( -S )
\label{Z}
\end{equation}
and evaluate this path integral within the saddle point approximation. The
least action paths 
\begin{equation}
\delta S /|\delta \Delta | =0, \;\;\;\;\;\;\delta S /\delta \varphi =0
\label{instantons}
\end{equation}
determine all possible QPS configurations. Integrating over small fluctuations
around all QPS trajectories one represents the grand partition function in
terms of infinite series (each term in such series corresponds to one
particular QPS saddle point). Then -- at least if interaction between
different quantum phase slips is small and can be neglected -- one can
easily sum these series and represent the final result in the form of the
exponent
\begin{equation}
Z=\exp (-F/T),
\label{ZF}
\end{equation}
where a formal expression for the free energy $F$ reads 
\begin{equation}
F=F_0-T 
\frac{\int{\cal D}\delta Y\exp(-\delta^2S_1[\delta Y])}{\int{\cal D}\delta Y\exp(-\delta^2S_0[\delta Y])}
\exp (-S_{QPS})\equiv F_0 -\frac{\gamma_{QPS}}{2}.
\label{freeinst}
\end{equation}
Here $F_0$ is the free energy without quantum phase slips, 
$\delta Y=(\delta\Delta,\delta\varphi)$ denote the fluctuations of relevant coordinates (fields), 
$\delta^2S_{0,1}[\delta Y]$ are the quadratic in $\delta Y$ parts of the action,
and the subscripts "0" and "1" denote
the action respectively without and with one QPS. 

The integrals over fluctuations in eq. (\ref{freeinst}) can be evaluated
exactly only in simple cases. Technically such a calculation can be quite
complicated even if the saddle point trajectories can be determined
explicitly. In our case an analytical expression for the QPS trajectory is not
even known. Hence, an exact evaluation of the path integrals in eq.
(\ref{freeinst}) is not possible.

Below we will present a simple approach which allows to establish 
the correct expression for the pre-exponent $B$ up to an unimportant numerical
prefactor. Within our present analysis any attempt to find an explicit value
for such a prefactor would make little sense simply because
the numerical value of $A$ in eq. (\ref{otvet}) is not known exactly.
Also for other problems numerical prefactors in the pre-exponent are usually
of little interest.  Therefore we believe that our approach may be useful for
various other situations because it allows to establish the correct functional
form of the pre-exponent practically without any calculation. If needed, with
a little extra effort our method may also allow to approximately evaluate a
numerical coefficient in the pre-exponent.

In order to calculate the ratio of the path integrals in eq. (\ref{freeinst})
let us introduce the basis in the functional space $\Psi_k(z)$ in which
the second  variation of the action around the instanton $\delta^2S_1[\delta
Y]$ is diagonal. Here the basis functions depend on a general vector
coordinate $z$ which is simply $z=(\tau,x)$ in our case. 
The first $N$ functions $\Psi_k$ are the so-called ``zero
modes'' related to the invariance of the instanton action under arbitrary
shifts in certain directions in the functional space (in our case -- shifts
of the QPS position along the wire and in imaginary time, i.e. $N=2$). Let
us denote an instanton solution as $\tilde Y(z)$. Then the zero mode
eigenfunctions are expressed as follows: $\Psi_k(X)=\partial \tilde Y/\partial
z_k$, where $k \leq N$ and the number of zero modes $N$ coincides with
the dimension of the vector $z$. An arbitrary fluctuation $\delta Y(z)$ can
be represented in terms of the Fourier expansion 
\begin{equation}
\delta Y(z)=\sum\limits_{k=1}^N \delta z_k \frac{\partial \tilde Y(z)}{\partial
z_k}+ \sum\limits_{k=N+1}^{\infty} u_k\Psi_k(z). 
\label{expan}
\end{equation}
Then we get
\begin{equation}
\delta^2S_0[\delta Y]=\frac{1}{2}\sum\limits_{k,n=1}^{\infty}
A_{kn}u_ku_n^*,\;\;\;\;\; \delta^2S_1[\delta
Y]=\frac{1}{2}\sum\limits_{k=N+1}^{\infty} \lambda_k |u_k|^2, 
\label{deS}
\end{equation}
where for $k\leq N$ the Fourier coefficients $u_k\equiv\delta z_k$ 
are just the shifts of the instanton position along the $k-$th axis
and $\lambda_k$ are the eigenvalues of $\delta^2S_1[\delta Y]$.  
Integrating over the Fourier coefficients one obtains
\begin{equation} 
\frac{\int{\cal D}\delta Y\exp(-\delta^2S_1[\delta
Y])}{\int{\cal D}\delta Y\exp(-\delta^2S_0[\delta Y])} = \int\limits_{0}^{L_1}
d\delta x_1 .. \int\limits_{0}^{L_N} d\delta x_N \sqrt{\frac{\det
A_{kn}}{(2\pi)^N\prod\limits_{k=N+1}^{\infty}\lambda_k}},  
\label{detex}
\end{equation}
where $L_k$ is the system size in the $k-$th dimension. The formula
(\ref{detex}) is, of course, not at all new. It just represents the standard
ratio of determinants with excluded zero modes \cite{ABC}. We will argue now,
that with a sufficient accuracy in the latter formula one can keep the
contribution of only first $N$ eigenvalues. Indeed, the contribution
of the ``fast''eigenmodes (corresponding to frequencies and
wave vectors much larger than the inverse instanton size in the corresponding
dimension) is insensitive to the presence of an instanton. Hence,
the corresponding eigenvalues are the same for both $\delta^2S_0$ and
$\delta^2S_1$ and just cancel out from eq. (\ref{detex}). In addition to the
fast modes there are several eigenmodes with frequencies (wave vectors) of
order of the inverse instanton size. The ratio between the product of all such
modes for $\delta^2S_1$ and the product of eigenvalues for $\delta^2S_0$
with the same numbers is dimensionless and  may only
affect a numerical prefactor which is not interesting for us here.
Dropping the contribution of all such eigenvalues one gets 
\begin{equation} 
\frac{\int{\cal D}\delta Y\exp(-\delta^2S_1[\delta
Y])}{\int{\cal D}\delta Y\exp(-\delta^2S_0[\delta Y])} \approx
\int\limits_{0}^{L_1} d\delta x_1 .. \int\limits_{0}^{L_N} d\delta x_N
\sqrt{\frac{\det A_{kn}|_{k,n\leq N}}{(2\pi)^N}}.  
\label{appz}
\end{equation} 
What remains is to estimate the parameters $A_{kk}$ for $k \leq N$.
For this purpose let us observe that the second variation of the action
becomes approximately equal to the instanton action,
$\delta^2S_1=\frac{1}{2}A_{kk}z_{0k}^2\approx S_{QPS}$, when the shift in
the $k-$th direction becomes equal to the instanton size in the same direction
$\delta z_k=z_{0k}$. Then we find $A_{kk}\approx 2S_{QPS}/z^2_{0k}$ and  
\begin{equation}
\det A_{kn}|_{k,n<N}\approx\prod\limits_{k=1}^N A_{kk}\approx
\frac{(2S_{QPS})^N}{\prod\limits_{k=1}^N z_{0k}^2}.
\label{prod22}
\end{equation} 
Finally, combining eqs. (\ref{gamma}), (\ref{freeinst}), (\ref{appz}) and
(\ref{prod22}) we obtain 
\begin{equation} 
B=bT\left(\prod\limits_{k=1}^N
\frac{L_k}{z_{0k}}\right) \left(\frac{S_{QPS}}{\pi}
\right)^{N/2}. 
\label{Ffinal} 
\end{equation} 
Here $b$ is an unimportant numerical prefactor. This result demonstrates that
the functional dependence of the pre-exponent can be determined
practically without any calculation. It is sufficient to know just the
instanton action, the number of the zero modes $N$ and the instanton
effective size $z_{0k}$ for each of these modes.

Let us also note that a similar observation has already been made \cite{ABC}
for some local Lagrangians equal to the
sum of kinetic and potential energies. Here we have shown that the
result (\ref{Ffinal}) holds for arbitrary effective actions, including nonlocal
ones. Hence, this result can be directly applied to our problem of quantum
phase slips  in thin superconducting wires. In this case we have $L_1\equiv
1/T$, $L_2 \equiv X$ and eq. (\ref{Ffinal}) yields 
\begin{equation}
B \approx \frac {S_{QPS}X}{\tau_0x_0}.
\label{B}
\end{equation}
This equation provides an accurate expression for the pre-exponent $B$
up to a numerical factor of order one. As we have already discussed, such an
accuracy is sufficient for our purposes. We also note that the result (\ref{B})
is parametrically different from previous results obtained within a TDGL-type of
analysis \cite{Chang} or suggested phenomenologically in Ref. \onlinecite{Gio}.

Finally, it is worth mentioning that we have also
compared our eq. (\ref{Ffinal}) with the exact results previously obtained for
various other problems by means of different approaches. Here we will briefly
discuss three different examples for the sake of illustration. The first
example is the problem of a quantum particle in a cosine periodic potential.
In this case an explicit expression for the inter-well tunneling rate is well
known. If we apply our method and evaluate only $A_{11}$, we will obtain the
tunneling rate which is $\sim 40$ \% smaller than the exact result. If
we also evaluate $A_{22}$ and $\lambda_2$ and include their ratio into our
formula, the result for the tunneling rate will be only 10 \% smaller
as compared to the exact one. This example demonstrates that also a sufficient
numerical accuracy  in the pre-exponent can be achieved without a complicated
calculation of the ratio of the determinants.

Two other examples concern the systems with nonlocal in time Lagrangians.
Consider, e.g., the problem of quantum decay of a particle in the presence
of dissipation \cite{cl}. In the limit of strong dissipation this problem
was treated by Larkin and Ovchinnikov \cite{LO} who found the exact
eigenvalues and, evaluating the ratio of the determinants, obtained the
prefactor in expression for the decay rate $B \propto \eta^{7/2}/m^2$,
where $\eta$ is an effective friction constant and $m$ is the particle mass.
This result would imply that the pre-exponential factor in the decay rate
\cite{LO} should be very large and may even diverge if one formally sets $m \to
0$.  

Later it was realized \cite{ZP} that this divergence is artificial.
Performing a simple one-loop perturbative calculation \cite{ZP} one 
arrives exactly at the same high frequency divergence as in the result
\cite{LO}. It implies that this divergence has nothing to do with tunneling,
and it is regularized by means of a proper renormalization of the bare
parameters in the effective action. After that the high frequency contribution
to the pre-exponent is eliminated and one finds \cite{ZP} $B \propto
1/\sqrt{\eta}$. [Note a misprint in the power of $\eta$ in eq. (8) 
of Ref. \onlinecite{ZP}.] This expression does not contain the particle mass
$m$ at all. It also allowed to fully resolve a discrepancy with the
experiments \cite{Lukens}. Note that the result \cite{ZP} can also be
expressed in the form $B \sim \sqrt{S_b}/\tau_0$, where $S_b \propto
\eta$ is the instanton (bounce) action and $\tau_0 \propto \eta$ is its
typical size. As we have already argued, the result in this form can be
guessed from eq. (\ref{Ffinal}) without any calculation. [For this 
particular problem our approach allows to even reproduce an exact numerical
prefactor.]

The last example is the problem of Coulomb blockade in normal tunnel junctions
in the strong tunneling limit. This problem was treated within the instanton
technique in Ref. \onlinecite{PZ}. In this problem each
instanton has two zero modes which correspond to its shifts in time and
fluctuations of its frequency $\Omega$. The value of the instanton
action is well known and the parameters $z_{01}$ and $z_{02}$ for both zero
modes can be evaluated directly by means of the approach presented above.
Each of these parameters is found to depend on one of the zero modes
$\Omega$. However, the product $z_{01}z_{02}$ turns out to be an
$\Omega$-independent constant. Making use of this fact 
and integrating over the zero mode coordinates in eq. (\ref{Ffinal}),
we arrive at the functional form of the pre-exponent derived in eq. (10) of
Ref. \onlinecite{PZ} by means of an explicit calculation of the fluctuation
determinants, see also Ref. \onlinecite{sz}.

The above examples demonstrate that our approach allows to
easily derive the functional form of the pre-exponent in a variety of
problems, including those where technically involved calculations appear to be
inevitable otherwise. 

\section{Comparison with Experiment}

Now let us compare our results with experimental findings. Recently
Bezryadin, Lau and Tinkham \cite{BT} reported a clear experimental evidence
for the existence of quantum phase slips in ultra-thin (with diameters down to
3 nm) and uniform in thickness  superconducting wires. Three out of eight
samples studied in the  experiments \cite{BT} showed no sign of 
superconductivity even well below the bulk critical temperature 
$T_C$. Furthermore, in the low
temperature limit the resistance of these samples was found
to show a slight upturn with decreasing $T$. In view of that
one can conjecture that these samples may actually become insulating 
at $T \to 0$. The resistance of other five samples \cite{BT} decreased
with decreasing $T$. Also for these five samples no 
clear superconducting phase transition was observed. 

All three non-superconducting wires (i1,i2 and i3) had the normal state
resistance below the quantum unit $R_q$,
while the normal state resistance of the remaining five ``superconducting'' 
samples was larger than $R_q$. This observation allowed the authors
\cite{BT} to suggest that a dramatic difference in the behavior of these
two group of samples (otherwise having similar parameters) can be due
to the dissipative phase transition (DPT) \cite{ZGOZ2,s,sz} analogous to that
observed earlier in Josephson junctions \cite{Mikko}. 

Without going into details here, let us just point out that DPT 
can be observed only provided quantum phase slips 
are easily created inside the wire. The results for $\gamma_{QPS}$
derived in the present paper allow to estimate a typical average time
within which one QPS event occurs in the sample. Making use of eqs.
(\ref{otvet}), (\ref{gamma}) and (\ref{B}) we performed an estimate of such
a time $t_0 = 1/\gamma_{QPS}$ for all eight samples studied in Ref. 
\onlinecite{BT}.
In this experiment the samples were fabricated from Mo$_{79}$Ge$_{21}$ alloy.
For our estimates we will use the value of the density of states 
$N_0=1.86\times 10^{13}$ sec/m$^3$ for clean Mo, which can be extracted 
from the specific heat data. The resistivity of the material
was measured to be $\rho=1.8$ $\mu\Omega/$m, the superconducting 
critical temperature is $T_C\simeq 5.5$ K. With these numbers we obtain 
the coherence length  $\xi \simeq 7$ nm in agreement with the estimate
\cite{BT}. The results for $t_0$ are summarized in the following Table: 
\begin{center}
\begin{tabular}{|c|c|c|c|c|}
\hline 
sample & $R/d$, k$\Omega$/nm & $S_{0}$ & $t_0|_{A=1},$ sec & $t_0|_{A=2},$ sec \\ 
\hline 
i1          & 0.122   &  7.8  &  $ 10^{-11}$   & $10^{-8}$ \\
\hline 
i2          & 0.110   &  8.7  &   $10^{-11}$   &  $10^{-6}$ \\ 
\hline 
i3          & 0.079   &  12.7  &   $ 10^{-9}$  &  $10^{-4}$ \\ 
\hline 
s1         & 0.038   &  25.1  &   $10^{-4}$   &  $10^{6}$ \\ 
\hline 
s2         & 0.028   &  33.7  &   $ 1$           &  $10^{14}$ \\ 
\hline 
s3         & 0.039   &  22.6  &   $ 10^{-5}$  &   $10^{5}$ \\ 
\hline 
ss1       & 0.054   &  15.4  &   $ 10^{-7}$  &  $10^{-1}$ \\ 
\hline 
ss2       & 0.044   &  19.6  &   $10^{-6}$   &  $10^{2}$ \\ 
\hline
\end{tabular}
\end{center}
The action $S_0$ is defined by  Eq. (\ref{otvet}) with $A=1$. 
The typical QPS time $t_0$ 
$$
t_0=\frac{\xi}{XAS_0 \Delta }\exp(AS_0).
$$
is very sensitive to the particular value of the factor $A$,
therefore here we present two estimates corresponding to $A=1$
and $A=2$.

In spite of remaining uncertainty in the prefactors some important
conclusions can be drawn already from the above estimates. For instance, 
we observe that for both $A=1$ and $A=2$ the QPS rate $\gamma_{QPS}=1/t_0$ is
very high (as compared, e.g., to the typical experimental time scale $\sim 1$
sec) in the ``insulating'' wires i1, i2 and i3. This fact is fully
consistent with the observations \cite{BT}: numerous quantum phase slips
occurring in these wires completely destroy the phase coherence and, hence,
superconductivity is washed out. Thus, non-superconducting behavior of these
three samples should be due to quantum phase slips.

On the other hand, the QPS rate
is notably lower for all the ``superconducting'' wires \cite{BT}. 
Possible interpretation of the experimental results for the samples
s1-ss2 depends strongly on the value of $A$.
E.g. for $A=1$ the QPS rate is high enough practically in all samples.
In this case quantum phase slips should in principle be important also
for ``superconducting'' wires \cite{BT}. Then one can indeed relate
the behavior of these samples to DPT \cite{ZGOZ2}, as a result of 
which quantum phase slips are bound in pairs and, hence, quantum 
fluctuations are strongly suppressed.

If, however, one chooses $A=2$ the QPS time for the samples
s1-ss2 turns out to be very long,
much longer than the experimental time. Then the QPS effects should
be irrelevant, and one would expect these samples to show a
superconducting behavior, perhaps with the renormalized critical temperature
\cite{Fin}. This conclusion  would also be consistent with the experimental
observations \cite{BT}.

Finally, let us note that all the above estimates are performed in the
limit $T=0$. This is correct if temperature is considerably below $T_C$. 
Otherwise the expression for the QPS action $S_{QPS}$ needs to be modified.

In conclusion, we have developed a detailed microscopic theory of
quantum phase slips in ultra-thin homogeneous superconducting wires.
We have derived the effective QPS rate for such wires and evaluated
this rate for the systems studied in recent experiments \cite{BT}. 
Our results are fully consistent with the experimental findings \cite{BT}  
which provide perhaps the first unambiguous evidence for 
QPS in mesoscopic metallic wires.

\appendix

\section{}

Let us collect some rigorous expressions for the 
``susceptibilities'' $\chi_E$, $\chi_J$, $\chi_L$ and $\chi_A$. In Ref. 
\onlinecite{ogzb} these quantities have been related to the so-called
polarization bubbles $f_0$, $g_0$ and $h_0$. In the interesting for us 
diffusive limit $\Delta l/v_F\ll 1$ these polarization bubbles are defined by
the equations \cite{ogzb}
\begin{equation}
f_0=T\sum\limits_{\omega_\nu}\int\frac{d^3{\bf k}}{(2\pi)^3}
\left\langle F(\omega+\omega_\nu,{\bf q}+{\bf k}) 
F(\omega_\nu,{\bf k})
\right\rangle_{\rm dis}
=\pi N_0T\sum\limits_{\omega _\nu }\frac{\Delta^2}{WW^{\prime
}(W+W^{\prime }+Dq^2)},  \label{f0}
\end{equation}
\begin{equation}
g_0=T\sum\limits_{\omega_\nu}\int\frac{d^3{\bf k}}{(2\pi)^3}
\left\langle G(\omega+\omega_\nu,{\bf q}+{\bf k}) 
G(\omega_\nu,{\bf k})
\right\rangle_{\rm dis}
=-N_0+\pi N_0T\sum\limits_{\omega _\nu }\frac{WW^{\prime }-\omega (\omega+
\omega_\nu)}{WW^{\prime }(W+W^{\prime }+Dq^2)}, 
\end{equation}
\begin{equation}
h_0=T\sum\limits_{\omega_\nu}\int\frac{d^3{\bf k}}{(2\pi)^3}
\left\langle G(\omega+\omega_\nu,{\bf q}+{\bf k}) 
\bar{G}(\omega_\nu,{\bf k})
\right\rangle_{\rm dis}
=-\pi N_0T\sum\limits_{\omega _\nu }\frac{WW^{\prime }+\omega (\omega+
\omega_\nu)}{WW^{\prime }(W+W^{\prime }+Dq^2)}. 
\end{equation}
In these equations we use the notations
\begin{equation}
W=\sqrt{\omega_\nu^2+\Delta^2},\; \;\;\;\; W'=\sqrt{(\omega_\nu+\omega)^2+\Delta^2},
\end{equation}
$\omega_\nu = \pi T(2\nu +1)$ and $\omega=2\pi Tn$,
where $\nu$ and $n$ are arbitrary integers, and $\langle \dots\rangle_{\rm
dis}$ implies disorder averaging.

Let us start from the function $\chi _E$ defined as
\begin{equation}
\chi _E=-\frac{2e^2}{q^2}(f_0+g_0). 
\end{equation}
At $T=0$ we find 
\begin{equation}
\chi _E=\left\{ 
\begin{array}{c}
\frac \sigma {\Delta}\frac 1{2\sqrt{1+x^2}}\int\limits_0^{+\infty }dt%
\frac{1-t^2+\sqrt{(t^2-1)^2+4t^2/(1+x^2)}}{\sqrt{(t^2-1)^2+4t^2/(1+x^2)}%
\left( 1+t^2+\sqrt{(t^2-1)^2+4t^2/(1+x^2)}\right) },\quad y=0, \\ 
\frac \sigma {\Delta}\left( \frac 1{2y}-\frac \pi {4y}+\frac{\ln \left( y+%
\sqrt{y^2-1}\right) }{2y\sqrt{y^2-1}}\right) ,\quad x=0,\quad y>1, \\ 
\frac \sigma {\Delta}\left( \frac 1{2y}-\frac \pi {4y}+\frac 1{y\sqrt{%
1-y^2}}\arctan \left( \sqrt{\frac{1-y}{1+y}}\right) \right) ,\quad x=0,\quad
y<1.
\end{array}
\right.   
\label{chiElim}
\end{equation}
Here and below we set $x=\omega /2\Delta$ and $y=Dq^2/2\Delta$.
In the limit $|\omega |\gg 2\Delta$ we obtain 
\begin{equation}
\chi _E\simeq \frac \sigma {|\omega |+Dq^2}.  
\label{chiEnm}
\end{equation}
At small $\omega$ and $q$ we get
\begin{equation}
\chi_E=\frac{\pi\sigma}{8\Delta}\left[ 1-\frac{3}{8}\left(\frac{\omega}{2\Delta}\right)^2
-\frac{8}{3\pi}\frac{Dq^2}{2\Delta} \right].
\end{equation}
It is also possible to evaluate $\chi_E$ at $\omega=0$, $q=0$ and arbitrary $T$:
\begin{equation}
\chi_E(0,0)=\frac{\pi\sigma}{8\Delta}\left(\tanh\frac{\Delta}{2T}
-\frac{\Delta}{2T\left( \cosh\frac{\Delta}{2T}\right)^2} \right).
\end{equation}
For $\Delta\ll T$ we find
\begin{equation}
\chi_E(0,0)=\frac{\pi\sigma\Delta^2}{96 T^3}.
\end{equation}
In this limit also a more general expression for arbitrary $\omega$ and $q$
can be obtained:
\begin{eqnarray}
\chi_E&=&\frac{\sigma}{|\omega|+Dq^2}+\frac{4\sigma\Delta^2 Dq^2}{(\omega^2-D^2q^4)^2}
\left[\Psi\left(\frac{1}{2}+\frac{|\omega|+Dq^2}{4\pi T}\right)-\Psi\left(\frac{1}{2}\right)\right]
\nonumber\\
&&
-\frac{2\sigma\Delta^2(\omega^2+D^2q^4)}{|\omega|(\omega^2-D^2q^4)^2}
\left[\Psi\left(\frac{1}{2}+\frac{|\omega|}{2\pi T}\right)-\Psi\left(\frac{1}{2}\right)\right]
+\frac{\sigma\Delta^2}{2\pi T(\omega^2-D^2q^4)}
\Psi'\left(\frac{1}{2}+\frac{|\omega|}{2\pi T}\right),
\end{eqnarray}
where $\Psi (x)$ is the digamma function.

The remaining $\chi$-functions are evaluated analogously. Consider the function
\begin{equation}
\chi _J=4e^2f_0.
\end{equation}
After straightforward algebra we obtain
\begin{equation}
\chi _J=\left\{ 
\begin{array}{c}
4e^2N_0\frac{\ln \left( |x|+\sqrt{1+x^2}\right) }{2|x|\sqrt{1+x^2}},\quad y=0,
\\ 
4e^2N_0\left( \frac \pi {4y}-\frac{\ln \left( y+\sqrt{y^2-1}\right) }{2y%
\sqrt{y^2-1}}\right) ,\quad x=0,\quad y>1, \\ 
4e^2N_0\left( \frac \pi {4y}-\frac 1{y\sqrt{1-y^2}}\arctan \left( \sqrt{%
\frac{1-y}{1+y}}\right) \right) ,\quad x=0,\quad y<1.
\end{array}
\right. 
\end{equation}
In the limit of low frequencies and wave vectors one has
\begin{equation}
\chi _J= 2e^2N_0\left[ 1-
\frac{2}{3}\left(\frac{\omega}{2\Delta }\right)^2-\frac{\pi}{4}\frac{ Dq^2}{2\Delta }\right].
\end{equation}
At $\omega=0,$ $q=0$ and for arbitrary $T$ we get
\begin{equation}
\chi_J=2\pi e^2N_0\Delta^2 T\sum\limits_{\omega_\nu}\frac{1}{(\omega_\nu^2+\Delta^2)^{3/2}}=
\left\{
\begin{array}{ll}
2e^2N_0, & T\ll\Delta, \\
\frac{7\zeta(3)}{2\pi^2}\,\frac{e^2N_0\Delta^2}{T^2}, & T\gg\Delta ,
\end{array}
\right.
\end{equation}
where $\zeta(3) \simeq 1.202$. 
In the limit $T\gg\Delta$ and $\omega\not=0$ we find
\begin{equation}
\chi_J=\frac{8e^2N_0\Delta^2}{|\omega|(\omega^2-D^2q^4)}\left\{
|\omega|\left[ \Psi\left(\frac{1}{2}+\frac{|\omega|+Dq^2}{4\pi T} \right) -\Psi\left(\frac{1}{2}\right)\right]
-Dq^2\left[ \Psi\left(\frac{1}{2}+\frac{|\omega|}{2\pi T} \right) -\Psi\left(\frac{1}{2}\right)\right]
\right\}.
\end{equation}

We proceed further with the function
\begin{equation}
\chi _L=\frac{8m^2\Delta^2}{q^2}\left( \frac 1\lambda +h_0-\left( 1+\frac{%
\omega ^2}{2\Delta^2}\right) f_0\right) .
\end{equation}
First we consider the limit $T=0$ and find
\begin{equation}
\chi _L=\left\{ 
\begin{array}{c}
4m^2\Delta N_0D\frac 1{\sqrt{1+x^2}}\int\limits_0^{+\infty }dt\frac{
1}{\sqrt{(t^2-1)^2+4t^2/(1+x^2)}},\quad y=0, \\ 
4m^2\Delta N_0D\frac 1{\sqrt{y^2-1}}\ln \left( y+\sqrt{y^2-1}\right)
,\quad x=0,\quad y>1, \\ 
4m^2\Delta N_0D\frac 2{\sqrt{1-y^2}}\arctan \left( \sqrt{\frac{1-y}{1+y}}%
\right) ,\quad x=0,\quad y<1.
\end{array}
\right. 
\end{equation}
At low frequencies and wave vectors the above expressions yield
\begin{equation}
\chi _L= 2\pi N_0Dm^2\Delta\left[ 1-
\frac{1}{4}\left(\frac{\omega}{2\Delta}\right)^2-\frac{2}{\pi}\frac{ Dq^2}{2\Delta}\right].
\end{equation}
For high temperatures $T\gg\Delta$ we obtain
\begin{equation}
\chi_L=\frac{4m^2\sigma\Delta^2}{e^2(\omega^2-D^2q^4)}\left\{
|\omega|\left[ \Psi\left(\frac{1}{2}+\frac{|\omega|}{2\pi T} \right) -\Psi\left(\frac{1}{2}\right)\right]
-Dq^2\left[ \Psi\left(\frac{1}{2}+\frac{|\omega|+Dq^2}{4\pi T} \right) -\Psi\left(\frac{1}{2}\right)\right]
\right\}.
\end{equation}
In the limit of zero frequency and wave vectors $\chi_L$ reduces to
a very simple form
\begin{equation}
\chi_L=\frac{\pi m^2\sigma\Delta^2}{2e^2 T}.
\end{equation}

Finally, let us evaluate the function
\begin{equation}
\chi _A=2\left( \frac 1\lambda +h_0+f_0\right) .
\end{equation}
We obtain
\begin{equation}
\chi _A=\left\{ 
\begin{array}{c}
2N_0\frac{\sqrt{1+x^2}}{|x|}\ln \left( |x|+\sqrt{1+x^2}\right) ,\quad y=0, \\ 
2N_0\left( \frac \pi {2y}+\frac{\sqrt{y^2-1}}y\ln \left( y+\sqrt{y^2-1}%
\right) \right) ,\quad x=0,\quad y>1, \\ 
2N_0\left( \frac \pi {2y}-\frac{2\sqrt{1-y^2}}y\arctan \left( \sqrt{\frac{1-y%
}{1+y}}\right) \right) ,\quad x=0,\quad y<1.
\end{array}
\right. 
\end{equation}
In the limit of high frequencies $|\omega |\gg \Delta$ one finds 
\begin{equation}
\chi _A\simeq 2N_0\ln\frac{|\omega|+Dq^2}{\Delta}.
\end{equation}
At low frequencies and wave vectors we derive
\begin{equation}
\chi _A\simeq 2N_0\left( 1+
\frac{1}{3}\left(\frac{\omega}{2\Delta}\right)^2+\frac{\pi}{4}\frac{ Dq^2}{2\Delta}\right) .
\end{equation}
In the high temperature limit $T\gg\Delta$ we find
\begin{equation}
\chi_A=2N_0\left[ \ln\frac{T}{T_C}+\Psi\left(\frac{1}{2}+\frac{|\omega|+Dq^2}{4\pi T} \right)
-\Psi\left(\frac{1}{2}\right)  \right].
\end{equation}

The above expressions are sufficient to evaluate the QPS action 
practically in all interesting limiting cases.


\begin{references}
\bibitem{almt}	L.G. Aslamazov, and A.I. Larkin,
		Fiz. Tverd. Tela {\bf 10}, 1140 (1968)
		[Sov. Phys. Solid State {\bf 10}, 875 (1968)];
		K. Maki, Prog. Theor. Phys. {\bf 39}, 897 (1968);
		R.S. Thompson, Phys. Rev. B {1}, 327 (1970).
\bibitem{HMW} P.C. Hohenberg, Phys. Rev. 158, 383 (1967);
  N.D. Mermin and H. Wagner, Phys. Rev. Lett. 17, 1133 (1966).
\bibitem{Little} W.A. Little, Phys. Rev. {\bf 156}, 396 (1967).
\bibitem{lamh}	J.S. Langer, and V. Ambegaokar, Phys. Rev. 
                {\bf 164}, 498 (1967). 		       
\bibitem{2}     D.E. McCumber, and B.I. Halperin,
		Phys. Rev. B {\bf 1}, 1054 (1970).
\bibitem{Exp}	J.E. Lukens, R.J. Warburton, and W.W. Webb, Phys. Rev.
Lett. {\bf 25}, 1180 (1970); R.S. Newbower, M.R. Beasley, and M. Tinkham, 
	Phys. Rev. B {\bf 5}, 864 (1972). 
\bibitem{Mooji} A.J. van Run, J. Romijn, and J.E. Mooij, Jpn. J. Appl. Phys.
{\bf 26} (1), 1765 (1987).
\bibitem{Gio}	N. Giordano, Phys. Rev B {\bf 43}, 160 (1991);
		{\it ibid.}, {\bf 41}, 6350 (1990);
		Physica B {\bf 203}, 460 (1994) and refs. therein.
\bibitem{Her}	F. Sharifi, A.V. Herzog, and R.C. Dynes,
                Phys. Rev. Lett. {\bf 71}, 428 (1993).
		A.V. Herzog, P. Xiong, F. Sharifi, and R.C. Dynes,
		Phys. Rev. Lett. {\bf 76}, 668 (1996);
                P. Xiong, A.V. Herzog, and R.C. Dynes,
                Phys. Rev. Lett. {\bf 78}, 927 (1997).
\bibitem{Lin}	X.S. Ling {\it et al.},
                Phys. Rev. Lett. {\bf 74}, 805 (1995).
\bibitem{sm}	S. Saito, and Y. Murayama, Phys. Lett. A {\bf 135}, 55 (1989);
		{\it ibid.} A {\bf 139}, 85 (1989).
\bibitem{Duan}	J.-M. Duan,
		Phys. Rev. Lett. {\bf 74}, 5128 (1995).
\bibitem{Chang} Y. Chang, Phys. Rev B {\bf 54}, 9436 (1996).
\bibitem{ZGOZ}  A.D. Zaikin, D.S. Golubev, A. van Otterlo,                 
                and G.T. Zimanyi, Phys. Rev. Lett. {78}, 1552 (1997).
\bibitem{ZGOZ2} A.D. Zaikin, D.S. Golubev, A. van Otterlo,             
                and G.T. Zimanyi, Usp. Fiz. Nauk {\bf 168}, 244 (1998)       
                [Physics Uspekhi {\bf 42}, 226 (1998)]. 
\bibitem{ogzb}  A. van Otterlo, D.S. Golubev, A.D. Zaikin, and G. Blatter,   
                Eur. Phys. J. B {\bf 10}, 131 (1999). 
\bibitem{BD}    R.M.Bradley and S. Doniach, Phys. Rev. B {\bf 30}, 1138 (1994).
\bibitem{many}  P.Bobbert, R.Fazio, G.Sch\"on, and A.D.Zaikin. Phys.          
      Rev. B {\bf 45}, 2294 (1992). \bibitem{HG}    F.W.J. Hekking and L.I.
Glazman, Phys. Rev. B {\bf 55}, 6551                 (1997). 
\bibitem{BT}    A. Bezryadin, C.N. Lau, and M. Tinkham, Nature {\bf
                404}, 971 (2000). 
\bibitem{s}     A. Schmid, Phys. Rev. Lett. {\bf 51}, 1506
                (1983); S.A. Bulgadaev, Zh. Eksp. Teor. Fiz. Pis'ma Red. 
                {\bf 39}, 264 (1984) [JETP Lett. {\bf 39}, 315 (1984)];     
                F. Guinea, V. Hakim, and A. Muramatsu, Phys. Rev. Lett.,     
                {\bf 54}, 263 (1985); M.P.A. Fisher and W. Zwerger, Phys. Rev.
                B {\bf 32}, 6190 (1985). 
\bibitem{sz}	G. Sch\"{o}n and A.D. Zaikin, Phys. Rep. {\bf 198},
                237 (1990). 
\bibitem{ms}	J.E. Mooij and G. Sch\"{o}n, Phys. Rev. Lett. {\bf 55}, 114 (1985). 
\bibitem{popov}	V.N. Popov, {\it Functional Integrals and Collective 
	        Excitations}, (Cambridge University Press, 1987).
\bibitem{klein}	H. Kleinert, Fortschr. Phys. {\bf 26}, 565 (1978).
\bibitem{ABC}   See, e.g., A.I. Vainstein, V.I. Zakharov, V.A. Novikov, and   
                M.A. Shifman, Usp. Fiz. Nauk {\bf 136}, 553 (1982) [Sov. Phys.
                Uspekhi {\bf 25}, 195 (1982)].  
\bibitem{Mikko} J.S. Penttila, P.J. Hakonen, M.A. Paalanen, and E.B.
                Sonin, Phys. Rev. Lett. {\bf 82}, 1004 (1999). 
\bibitem{Fin}   Y. Oreg and A.M. Finkelstein, Phys. Rev. Lett. {\bf 83}, 191
                (1999). 
\bibitem{cl}    A.O. Caldeira, and A.J. Leggett,
		Phys. Rev. Lett. {\bf 46}, 211 (1981);
		Ann. Phys. (N.Y.) {\bf 149}, 347 (1983).
\bibitem{LO}    A.I. Larkin and Yu.N. Ovchinnikov, Zh. Eksp. Teor. Fiz. 
                {\bf 86}, 719 (1984) [Sov. Phys. JETP {\bf 59}, 420 (1984)].
\bibitem{ZP}    A.D. Zaikin and S.V. Panyukov, Zh. Eksp. Teor. Fiz. Pis'ma Red. 
                {\bf 43}, 518 (1986) [JETP Lett. {\bf 43}, 670 (1986)].
\bibitem{Lukens} D.B. Schwartz, B. Sen, C.N. Archie, and J.E. Lukens, Phys.
                 Rev. Lett. {\bf 55}, 1547 (1985). 
\bibitem{PZ}  S.V. Panyukov and A.D. Zaikin, Phys. Rev. Lett. {\bf 67},
                3168 (1991).
\end{references}
\end{document}